\documentclass[a4paper,11pt]{article}
\pdfoutput=1
\usepackage{jinstpub}
\usepackage{lineno}
\usepackage[per-mode=symbol, separate-uncertainty=true]{siunitx}
\usepackage{xspace}
\usepackage{subcaption}
\usepackage{booktabs}
\usepackage{multirow}
\usepackage[noabbrev,capitalise]{cleveref}

\DeclareSIUnit[]\electron{\elementarycharge ^{-}}

\newcommand*{\TeV}{\ensuremath{\text{Te\kern -0.1em V}}}
\newcommand*{\GeV}{\ensuremath{\text{Ge\kern -0.1em V}}}
\newcommand*{\MeV}{\ensuremath{\text{Me\kern -0.1em V}}}
\newcommand*{\ttbar}{\ensuremath{t\bar{t}}\xspace}

\newcommand*{\Ntrkevent}{\ensuremath{N_\text{trk}^\text{event}}\xspace}
\newcommand*{\Ntrkwrong}{\ensuremath{N_\text{trk}^\text{wrong}}\xspace}

\newcommand*{\Ntrk}{\ensuremath{N_\text{trk}}\xspace}
\newcommand*{\Nclntrkone}{\ensuremath{N_\text{ntrk{=}1}^\text{clust}}\xspace}
\newcommand*{\Nclntrkmany}{\ensuremath{N_\text{ntrk{>}1}^\text{clust}}\xspace}
\newcommand*{\Nvrtntrkone}{\ensuremath{N_\text{ntrk=1}^\text{vrt}}\xspace}
\newcommand*{\Nvrtntrkmany}{\ensuremath{N_\text{ntrk>1}^\text{vrt}}\xspace}

\newcommand{\expect}[1]{\ensuremath{\bigl \langle #1 \bigr \rangle}}

\DeclareSIUnit\clight{\text{\ensuremath{c}}}
\DeclareSIUnit[per-mode=symbol]\GeVc{\GeV\per\clight}

\title{Improving primary-vertex reconstruction with a minimum-cost lifted multicut graph partitioning algorithm}
\date{\today}

\author[1]{V.~Kostyukhin,}
\author[2,3]{M.~Keuper,}
\author[1]{I.~Ibragimov,}
\author[1]{N.~Owtscharenko,}
\author[1]{and M.~Cristinziani}

\affiliation[1]{Center for Particle Physics Siegen, Department Physik, Universität Siegen, Walter-Flex-Straße 3, 57072 Siegen, Germany}
\affiliation[2]{Department Elektrotechnik und Informatik, Universität Siegen, Hölderlinstraße 3, 57076 Siegen, Germany}
\affiliation[3]{Max Planck Institute for Informatics, Saarland Informatics Campus E1 4, 66123 Saarbrücken, Germany}

\abstract{%
Particle physics experiments often require the simultaneous reconstruction of many interaction vertices. Usually, this problem is solved by ad hoc heuristic algorithms. 
We propose a universal approach to address the multiple vertex finding through a principled formulation as a minimum-cost lifted multicut problem. The suggested algorithm is tested in a typical LHC environment with multiple proton--proton interaction vertices. 
Reconstruction errors caused by the particle detectors complicate the solution and require the introduction of special metrics to assess the vertex-finding performance. We demonstrate that the minimum-cost lifted multicut approach outperforms heuristic algorithms and works well up to the highest vertex multiplicity expected at the LHC.
}

\arxivnumber{2301.05548}

\begin{document}
\maketitle

\section{Introduction}
\label{sec:intro}

In particle physics experiments, many problems require a precise reconstruction of vertices --- points in 3D space where particle interactions occur. Knowledge of the positions and features of such vertices provides valuable information about the underlying physics of these interactions. There are numerous examples: $B$ physics, heavy-flavour jet identification, primary event vertex reconstruction, search for exotic particles as new physics manifestations, etc.
Except for a few specifically designed detectors (emulsions, Wilson chamber, etc.), rarely used in modern experiments, the vertices are not directly detectable. The presence of vertices is usually inferred from 3D traces of charged stable or quasi-stable particles produced in the interaction. Various tracking detectors measure the trajectories of these particles (tracks) in space. The trajectories can be extrapolated to a single 3D point, which represents the interaction vertex position~\cite{Book_vertex}.

Despite the simplicity of the vertex reconstruction idea, its real-life exploitation encounters problems. For example, at the Large Hadron Collider (LHC) at the end of Run 2, a typical recorded event may exhibit up to $\sim$80 primary proton--proton interactions, and numerous produced charged particles underwent further interactions leading to additional vertices, distributed in significant 3D volumes. The expected number of proton--proton interactions in a single event at the LHC after the planned high-luminosity upgrade (HL-LHC) may reach 200--300, resulting in a few thousand reconstructed tracks.
The exact amount of vertices in this set of reconstructed tracks is not known a priori. Thus, an efficient inclusive vertex-reconstruction algorithm needs first to determine how many vertices are present in a given track collection, then to assign the reconstructed tracks to these assumed vertices and, finally, to determine the coordinates of each vertex.  The track measurement uncertainties, which may differ by a factor of 10 for different tracks and often are comparable with the vertex--vertex distances, cause additional complications. These uncertainties make an exact crossing of track pairs in 3D space impossible: even if two charged particles are produced in the same interaction point, their reconstructed trajectories will only be close to the true vertex position and to each other, up to the corresponding uncertainties.

 The explicit reconstruction of multiple vertices from a given track collection can be addressed in a graph-based approach. In fact, all space trajectories of the particles produced in a single vertex should be pairwise compatible, i.e.~every pair of tracks should be close to each other in some volume around the true vertex position. Therefore, a compatibility (adjacency) graph can be constructed where every node represents a track. Two nodes are connected by an edge if and only if the distance of the corresponding trajectories is small. In the ideal case, every vertex is represented by a fully connected, isolated subgraph in such a graph. In a realistic scenario, track measurement errors shuffle tracks among different vertices, resulting in a large number of fake edges in the compatibility graph. Yet, it can be tried to partition the full graph into non-overlapping components by cutting some edges so that the remaining edges represent track pairs with minimal distances. Each such graph component determines a union of close-by tracks which can be considered as a vertex approximation.    

The present paper focuses on finding primary proton--proton interaction vertices at the LHC. The typical transverse width of the LHC proton beam at the interaction region is $\sim$10$\,\mu$m, and the two beams cross each other at a very small angle. As a result, the primary interaction vertices occur in a very narrow and long volume which can be approximated by a line, making this problem effectively one-dimensional. Subsequent decays of the particles produced in the detector volume and their interactions with the detector material, requiring an explicit 3D treatment,  will not be considered here. Constraining the vertex positions to a line preserves all important data features while making data generation and comparison with existing approaches much easier. 

\begin{figure}[htbp]
\begin{center}
   \includegraphics[width=0.95\textwidth]{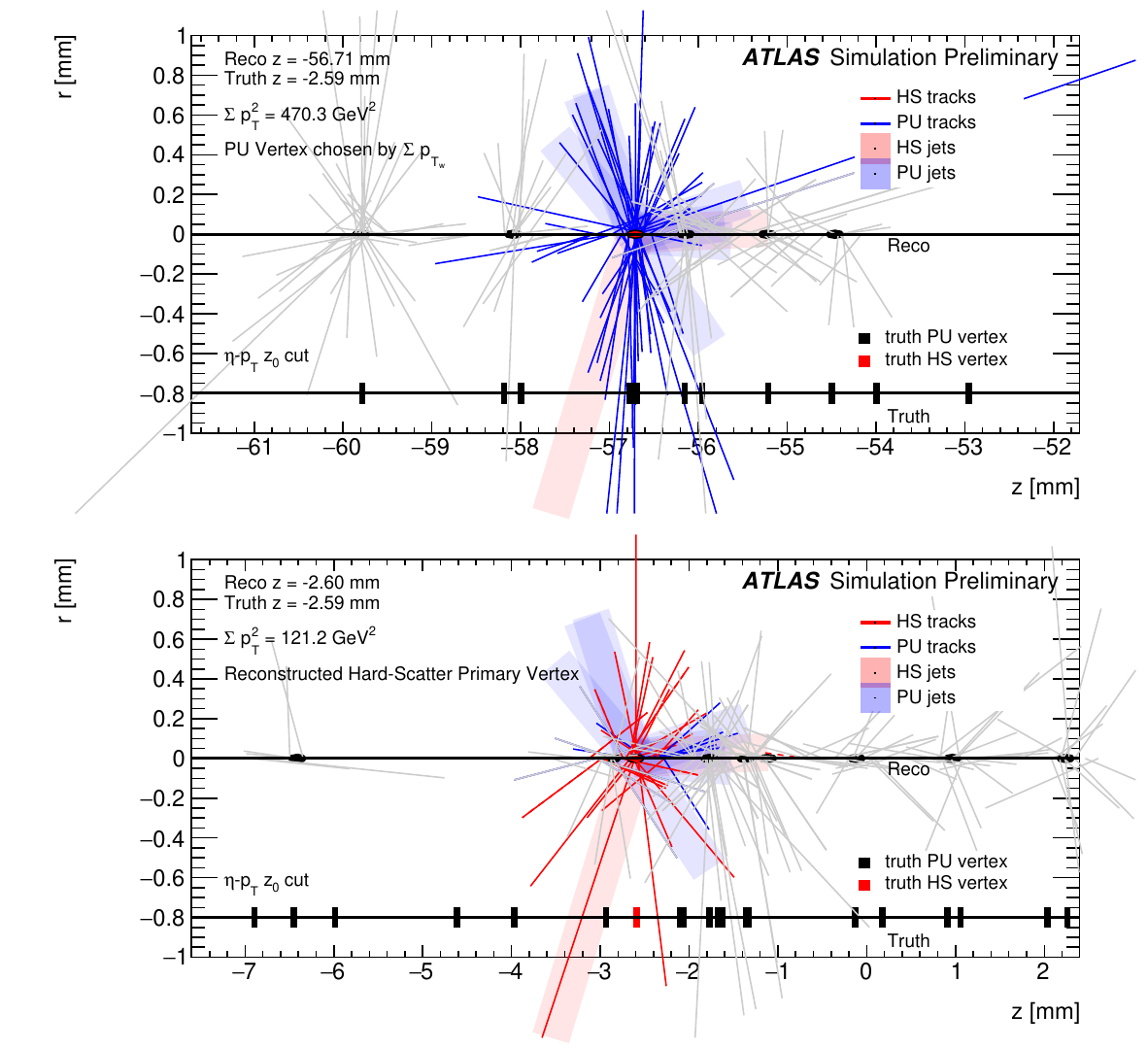}
   \caption{\label{fig:Event} Two regions of a typical LHC event in the ATLAS detector with many pileup interactions~\cite{IDTR-2019-004}. True positions of the proton--proton interactions are shown, as well as the reconstructed trajectories (tracks) of the produced particles scattered due to reconstruction uncertainties. Some truth interaction vertices do not have associated tracks because all emanated particles are outside of the sensitive detector phase space and not reconstructed. 
   These pictures illustrate typical track densities and overlap of the tracks produced in nearby interaction vertices. Both, tracks associated with the hard-scattering (HS) and pileup (PU) are shown. 
   }
\end{center}  
\end{figure}

To illustrate the primary-vertex reconstruction problem, \cref{fig:Event} shows two zoomed-in regions of a typical simulated LHC event with several pileup interactions.In both plots, the true positions of interaction vertices are shown, together with charged particle trajectories displaced due to reconstruction uncertainties. Several true interaction vertices in these plots do not have associated tracks because all emanated particles in this interaction are outside of the tracking detector's sensitive volume, see \cref{sec:Delphes} for the details. The overlap of the red (from hard-scatter vertex), blue and grey (from nearby pileup vertices) tracks in the centre of the bottom plot on \cref{fig:Event} is clearly visible.

Experiments at the LHC use heuristic algorithms~\cite{IVF, PUB-2019-015, CMStracker2014} to reconstruct multiple proton--proton interaction vertices.
Several other approaches can be found in the literature, including medical imaging-inspired algorithms~\cite{Hageb_ck_2012} and the RAVE package~\cite{RAVE} implementing the deterministic annealing algorithm~\cite{Annealing}. The latter targets a universal multi-vertex reconstruction, but was not yet used to directly reconstruct primary vertices.

This article presents an implementation of the Lifted Multicut Graph Partitioning algorithm (LMC), which solves the inclusive vertex reconstruction problem described above. \cref{sec:LMP} describes the LMC algorithm and details of its implementation for the vertex finding application.
 \cref{sec:Delphes} describes the simulated samples which are used to test the algorithm performance. In \cref{sec:truthfeatures}, features of the simulated samples are discussed. \cref{sec:wgtedge} introduces edge cost functions used in the graph partitioning. In \cref{sec:metrics}, the metrics are introduced to estimate the algorithm performance and to compare it with other existing approaches. 
 \cref{sec:results} presents the performance of the LMC approach
 in simulation. In \cref{sec:conclusions}, conclusions are made.

\section{Minimum-cost multicuts and lifted multicut algorithm for cluster finding}
\label{sec:LMP}

The compatibility (adjacency) graph representation of the reconstructed track set allows to formulate the primary-vertex finding problem as a minimum-cost lifted multicut graph-partitioning problem. This problem was originally proposed in Reference~\cite{Keuper_2015_ICCVa} in the context of image segmentation and mesh decomposition. It is a generalization of the better-known minimum cost multicut problem, also referred to as the weighted correlation clustering problem \cite{demaine-2006,chopra-1993}. The minimum cost multicut problem is a grouping problem defined for a graph $G=(V,E)$ and a cost function $c: E \rightarrow \mathbb{R}$ which assigns to all edges $e \in E$ a real-valued cost or reward for being cut. Then, the minimum cost multicut problem is to find a binary edge labelling $y$ according to 
\begin{align}
\min\limits_{y \in \{0, 1\}^{E}}
\sum\limits_{e \in E} c_e y_e
\label{eq:LMC1}
\end{align}
subject to
\begin{align}
\forall C \in \mathrm{cycles}(G) \quad \forall e \in C : y_{e} \leq \sum\limits_{e' \in C\backslash\{e\}} y_{e'}\,.
\label{eq:LMC2}
\end{align}
The constraints on the feasible set of labellings $y$ given in \cref{eq:LMC2} ensure that the solution of the multicut problem relates one-to-one to the decompositions of graph $G$, by ensuring for every cycle in $G$ that if an edge is cut within the cycle ($y_e=1$), so needs to be at least one other. Trivial optimal solutions are avoided by assigning positive (attractive) costs $c_e$ to edges between nodes $v,w \in V$ that likely belong to the same component, while negative (repulsive) costs are assigned to edges that likely belong to different components.

The minimum cost \emph{lifted} multicut problem (LMC) generalizes over the problem defined in \cref{eq:LMC1}--\cref{eq:LMC2} by adding a second set of edges that defines additional, potentially long-range costs without altering the set of feasible solutions. It thus defines a second set of edges $F$ between the nodes $V$ of $G$, resulting in a lifted graph $G'=(V,E\cup F)$, on which we can define a cost function $c':E\cup F \rightarrow \mathbb{R}$. Then, \cref{eq:LMC1} and \cref{eq:LMC2} are optimized over all edges in $E\cup F$ and two additional sets of constraints are defined according to \cite{Keuper_2015_ICCVa} 
\begin{align}
\forall v,w \in F \quad \forall P \in v,w-\mathrm{paths}(G) : y_{vw} \leq \sum\limits_{e\in P} y_{e}
\label{eq:LMC3}
\end{align}
\begin{align}
\forall v,w \in F \quad \forall C \in v,w-\mathrm{cuts}(G) : 1 - y_{vw} \leq \sum\limits_{e\in C} (1 - y_e)
\label{eq:LMC4}
\end{align}
to ensure that the feasible solutions to the LMC problem still relate one-to-one to the decompositions of the original graph $G$.

For the vertex-finding problem, this formulation allows encoding Euclidean distance constraints in the structure of graph $G$ (e.g.~point observations that are spatially distant can not originate from the same vertex), while the cost function can be naturally defined in the distance significance space to take into account the measurement errors. The Euclidean distance and its significance can be very different in case of significant reconstruction errors, the {\it lifted} multicut  formulation encodes both metrics in the same graph.

The minimum cost multicut problem is $np$-hard, and so is the minimum cost LMC problem \cite{2017PolytopeFigure}. Yet, efficient heuristic solvers provide practically good solutions \cite{CGC, 2015InferenceTechniques,Beier2016,Keuper_2015_ICCVa,amir_solving}. Here, we resolve to use the primal feasible heuristic KLj that has been proposed in Reference~\cite{Keuper_2015_ICCVa} and published in an open-source library\footnote{https://github.com/bjoern-andres/graph}. KLj is an iterative approach that produces a sequence of feasible solutions whose cost decreases monotonically. It takes as input an initial edge labelling (for example, all edge labels are initially set to $0$), a lifted graph and costs defined on all edges. In every step, it either moves nodes between two neighbouring components, moves nodes from one component into a new component or joins two components such as to decrease the cost of the multicut maximally according to \cref{eq:LMC1}. 

Finally, the compatibility (adjacency) graph partitioning defines the splitting of the initial track collection into a set of clusters consisting of linked tracks with minimal cost. This set of track clusters is a solution of the primary-vertex finding problem, each cluster representing a vertex from which all included tracks emanate.

\section{Data simulation}
\label{sec:Delphes}

To estimate the track clustering performance based on the compatibility graph partitioning with the LMC algorithm, we simulated data using DELPHES~\cite{delphes}. The framework allows to perform a fast and realistic simulation of a general-purpose collider detector composed of an inner tracker, electromagnetic and hadron calorimeters, and a muon system. For this study, we added a detailed parameterisation of the ATLAS detector tracking resolution to the framework.

To simulate the pileup vertices and hard-scattering events, a sufficiently large amount of minimum-bias interaction events was prepared, consisting of single, double, and non-diffractive processes. These events have been generated using the Pythia 8~\cite{pythia} event generator. As the main source of hard-scattering interactions, \ttbar events are used, also generated with Pythia 8. To simulate an LHC collision event with full pileup, a single \ttbar event is mixed with a number of minimum-bias events, distributed according to a Poisson distribution with a mean corresponding to a chosen luminosity. The interaction vertices are then distributed along the LHC beam trajectory inside the detector, according to typical interaction region parameters for ATLAS. Two different descriptions are used: a Gaussian with $\sigma_z=35$ mm for a collision energy of 13~TeV to be compared ATLAS Run~2 results~\cite{PUB-2019-015}, and a Gaussian with $\sigma_z=42$ mm for 14~TeV to emulate the HL-LHC environment~\cite{HLLHCperf}. Since the LHC beam width is very small, in this simulation it is neglected. 

The acceptance of the ATLAS detector allows for reconstructing charged particle trajectories in a limited phase space of $p_{\perp} > 500~\MeV$ and $|\eta|<2.5$. Some minimum-bias proton--proton interactions produce only particles outside the sensitive phase space of the ATLAS detector, which makes them unreconstructable. Positions of interactions with a single track in the ATLAS acceptance can be reconstructed, but this vertex category is contaminated by tracks that are strongly displaced by measurement errors. 
In the following, a {\it reconstructable truth vertex} refers to the true position of a proton--proton interaction producing at least two tracks within the ATLAS detector acceptance. 

All tracks produced in an event and falling into the sensitive ATLAS detector phase space are smeared according to the parameterised ATLAS detector resolution \cite{IBLTDR,IDTR-2016-018}. Tracks with smeared parameters are referred to as reconstructed tracks in the following. The set of reconstructed tracks corresponding to a full pileup event is used as input for the performance estimation of the clustering algorithms. DELPHES samples used in this paper have been
prepared with different energies and different pileup 
conditions~(\cref{tab:DataSamples}).

 \begin{table}[htb]
  \centering
   \begin{tabular}{ccccccc}
    \toprule
     Energy & \expect{\mu} & Interaction region $\sigma_z$ & 
     \expect{\Ntrkevent} & \expect{N_\text{trk=0}^\text{vrt} } & \
     \expect{N_\text{trk=1}^\text{vrt}}  & \expect{N_\text{trk>1}^\text{vrt}} \\
     \midrule
     13~TeV & 63 & 35 mm & 718 & 9  & 4 & 50 \\
     14~TeV & 150 & 42 mm & 1674 & 22  & 9 & 119 \\
     14~TeV & 200 & 42 mm & 2227 & 28  & 12 & 160 \\
     14~TeV & 250 & 42 mm & 2771 & 35  & 16 & 199 \\
     \bottomrule
  \end{tabular}
  \caption{DELPHES samples used to estimate the LMC performance. The column \Ntrkevent reports the total number of reconstructed tracks in simulated events. The last three columns show the average numbers of true vertices with $N_\text{trk}=0,1,{>}1$ correspondingly. True vertices with $N_\text{trk}{>}1$ are considered {\it reconstructable}.  \label{tab:DataSamples} }
\end{table}

\section{Features of simulated data} 
\label{sec:truthfeatures}

\begin{figure}[htbp]
  \centering
   \begin{tabular}{cc}    
      \includegraphics[width=0.47\textwidth]{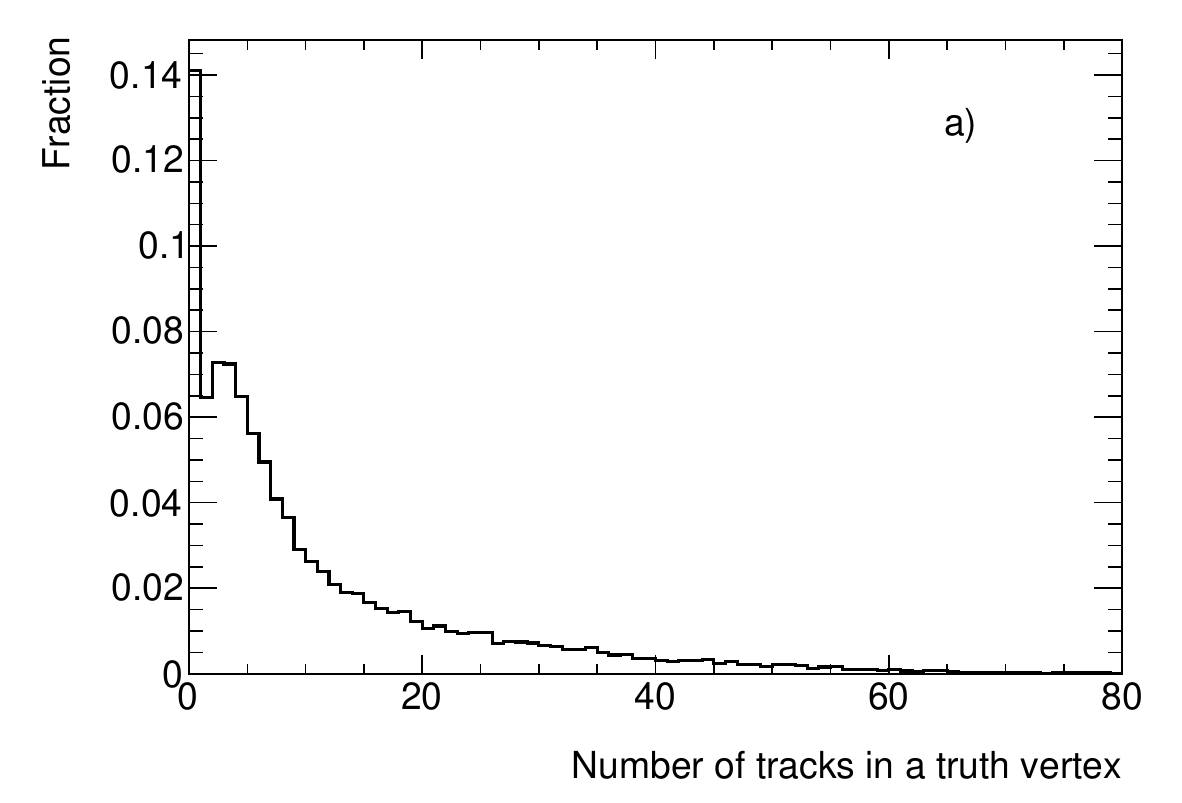}& 
      \includegraphics[width=0.47\textwidth]{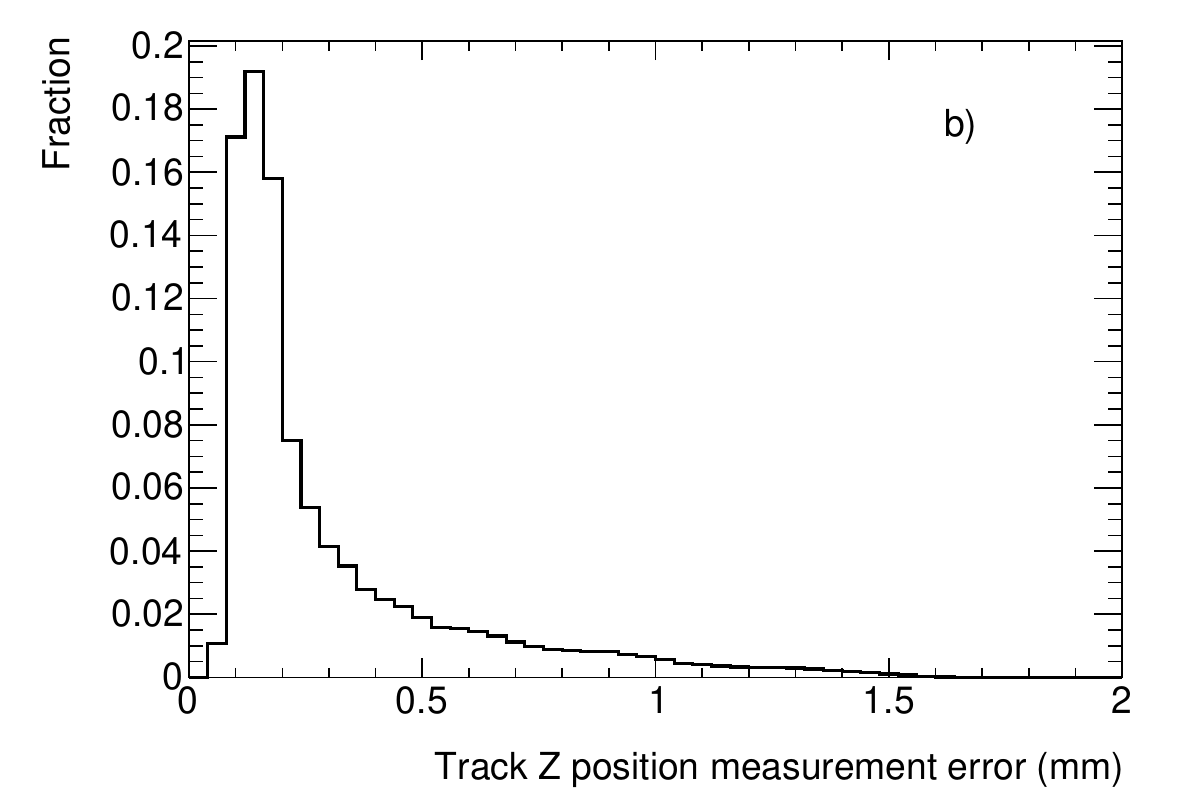}  \\
   \end{tabular} 
   \caption{ a) Number of tracks per simulated vertex and 
   b) longitudinal position measurement errors for simulated tracks at a collision energy of 13~TeV\label{fig:TruthVertex}.}
\end{figure}

The number of truth tracks in the detector acceptance in the simulated vertices and the position measurement errors of these tracks are shown in \cref{fig:TruthVertex} for a collision energy of 13~TeV. As can be seen in \cref{fig:TruthVertex}a, in 14\% of the cases, the simulated vertices do not have tracks in the detector acceptance, and in $6.5\%$ of the cases, they have only one track. The number of tracks for all other vertices is widely spread up to 80.  

\begin{figure}[htb]
  \centering
    \includegraphics[width=0.9\textwidth]{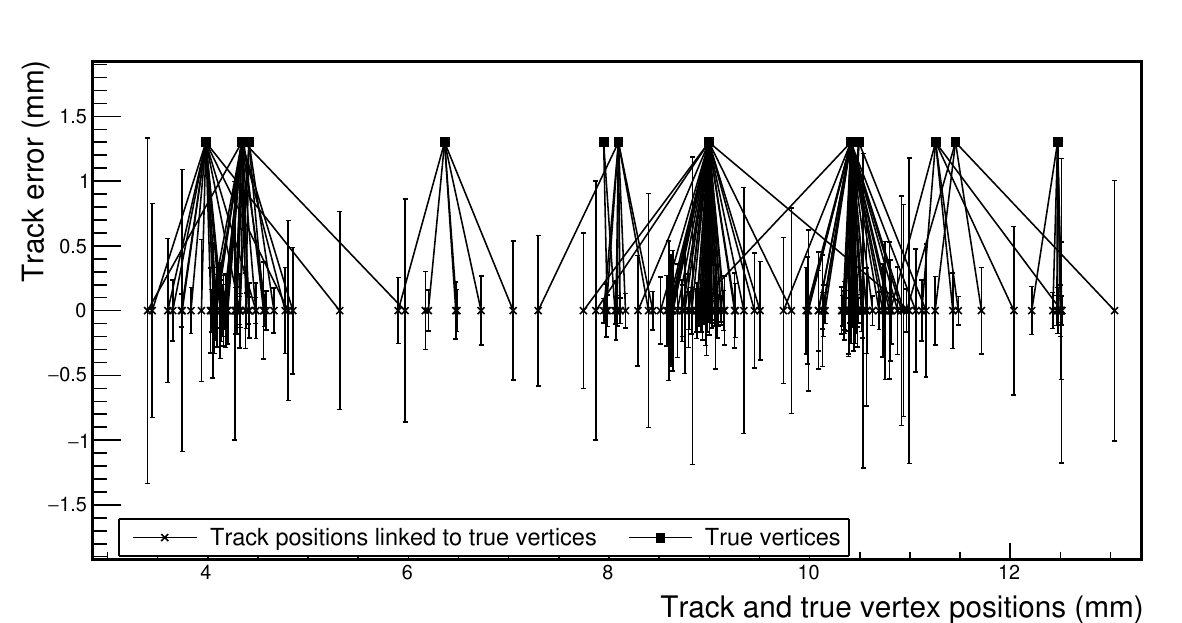}
  \caption{Example display of overlapping tracks from different vertices caused by measurement errors (zoom of a simulated DELPHES event with $\mu=150$). The crosses at the ordinate value of 0 represent the track positions, and the vertical error bars represent the corresponding position measurement errors. Squares at ordinate values of 1.3 represent the truth vertex positions. The connecting lines show the origin vertex for every track. \label{fig:overlap}}
\end{figure}

  Track measurement errors are shown in \cref{fig:TruthVertex}b. From the sizes of the luminous regions and the number of vertices in \cref{tab:DataSamples} we can conclude that the track measurement errors are 
  comparable or larger than a typical vertex--vertex distance in the simulated data.
  Smearing of the track positions due to measurement errors results in a significant overlap of the tracks from different truth vertices. This effect can be characterised by the truth track overlap fraction, i.e.~the fraction of reconstructed tracks in an event that are closer in space to another truth vertex than to the truth vertex of origin. This overlap fraction is shown in \cref{tab:TruthOverlap} as a function of pileup.
  An example of the track overlap can be seen in the bottom panel of \cref{fig:Event}. Another example is shown in \cref{fig:overlap}.

\begin{table}[htb]
  \centering
   \begin{tabular}{ccccc}
    \toprule
     \expect{\mu} & 63 & 150  & 200 & 250 \\
     \midrule
     Truth track overlap fraction & 20\% & 41\% & 53\%  &  66\% \\
     \bottomrule
  \end{tabular}
  \caption{ Average truth track overlap fraction in an event, as a function of pileup conditions.   \label{tab:TruthOverlap} }
\end{table}

  A priori, well-measured tracks with small errors should be easy to cluster according to the truth, while poorly measured tracks with large errors can easily migrate from one cluster to another, independently of their true origin. This random migration can be interpreted as noise, and thus, the overall problem may be considered as clustering in the presence of significant noise. 

\section{Edge weights and constraints}
\label{sec:wgtedge}
  
To formulate the vertex finding problem in the presence of pileup as a minimum cost lifted multicut (LMC) graph-partitioning problem, a track-pair compatibility graph needs to be constructed. A node in this graph represents a track, and two nodes are connected by an edge if and only if they are close in space and can be produced in the same vertex. The degree of track closeness, or equivalently the probability of originating from the same vertex, is estimated during the graph construction and is expressed as a weight assigned to the edge. The edge weights determine the efficiency of the partitioning. Therefore, they should incorporate enough information, and the weight assignment procedure should be carefully designed. The following approaches are used in our study:

\begin{enumerate}
\item Probability density function (PDF) ratio of the track--track geometrical distance significance based on measured uncertainties, $S = \sqrt{(z_i-z_j)^2 / (\sigma_i^2+\sigma_j^2)}$;
\item Multivariate binary classification with Boosted Decision Trees (BDT);
\item Logistic regression based on $S$.
\end{enumerate}

\begin{figure}[htbp]
  \centering
    \includegraphics[width=\textwidth]{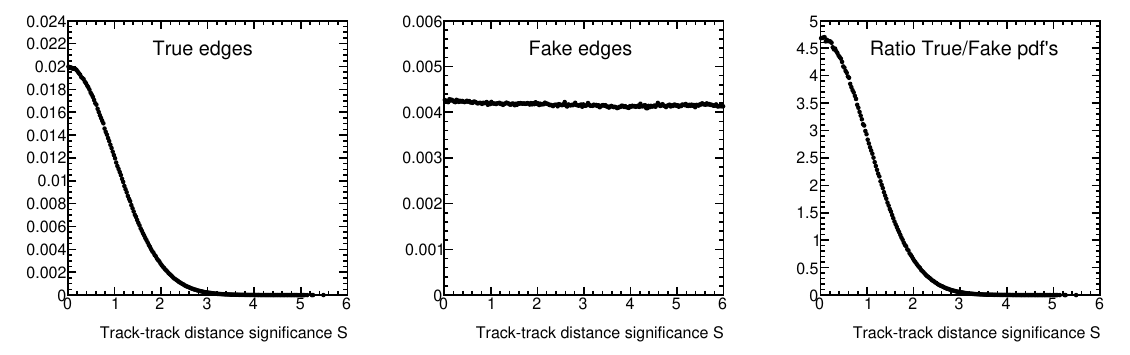} 
  \caption{Example track--track distance significance for true and fake edges and their ratio. The significance distributions are normalized to one.}
  \label{fig:EdgeWgt}
\end{figure}

The LMC formulation assumes that the correct edges (two tracks from the same vertex) receive positive weights, while random (fake) edges receive negative weights. This can be achieved by using a logarithm of the ratio of the probability density functions for the correct and fake edges as the cost function of the problem $\log{\frac{p_\mathrm{true}}{p_\mathrm{fake}}}$. According to the Neyman--Pearson lemma, this is the most efficient test statistic for the true/fake edge classification. An example of the track--track distance significance distributions and their ratio are shown in \cref{fig:EdgeWgt}. As the PDF of the fake edges is independent of the track--track distance significance, its overall normalisation depends on the significance range used for the parameterisation. Thus, the exact values of the PDF ratio can be scaled by the choice of the parametrisation range, which in principle, should not affect the LMC clustering performance if the range is sufficiently large. Such a behaviour can be mimicked by a global multiplier of the PDF ratio function. The influence of 
this multiplier on the clustering will be studied in \cref{subsec:stability}.

\begin{figure}[htbp]
  \centering
    \includegraphics[width=0.5\textwidth]{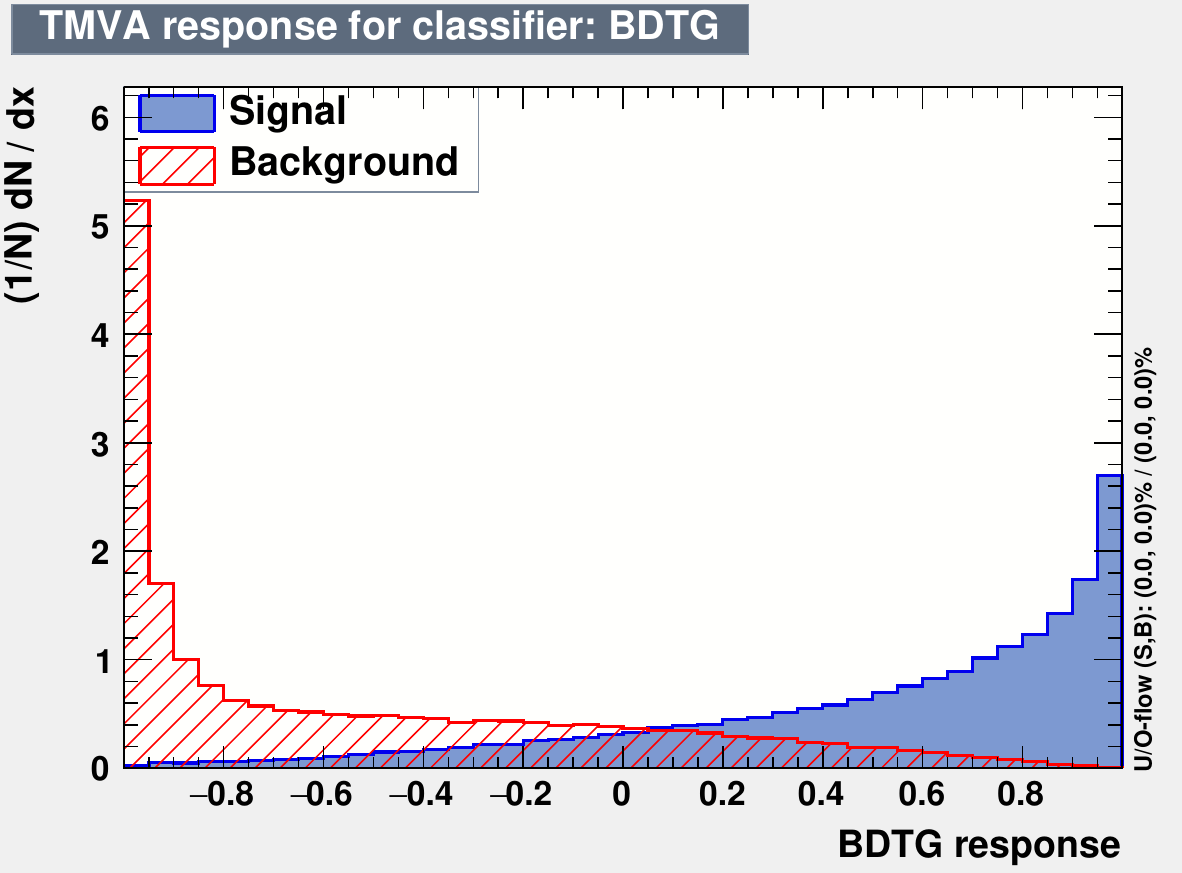} 
  \caption{Example BDTG classification weight distributions for true and 
         fake edges.}
  \label{fig:EdgeBDTG}
\end{figure}

A better clustering performance could be achieved by encoding more information in the edge weight calculation. To test this approach, we use a BDT classifier combining seven features, listed in \cref{tab:BDTvars}, to distinguish true edges from fake ones. The GradientBoost implementation (BDTG) from the TMVA~\cite{TMVA} package is used to train the classifier. An example of the trained classifier response\footnote{ TMVA GradientBoost uses the binomial log-likelihood loss $L(F,y)=\ln [1+\exp(-2F(x)y)]$ with Gini Index separation. We use the following training settings NTree=800, MaxDepth=10, MinNodeSize=1.5\%, Shrinkage=0.07. } is shown in \cref{fig:EdgeBDTG}.  The output is negative for fake edges and positive for true ones, exactly as required by the KLj algorithm, and therefore can be used directly as the edge weight.  

\begin{table}[htbp]
  \centering
   \begin{tabular}{cl}
    \toprule
   n. & Description \\
   \midrule
     1 & Squared significance $S^2$ (or $\chi^2$) of track--track distance along beamline \\
     2 & Average position of the track pair along beamline \\
     3 & Position measurement uncertainty $\sigma_{z_0}$ of  track 1 \\
     4 & Position measurement uncertainty $\sigma_{z_0}$ of  track 2 \\
     5 & Pseudorapidity $\eta$ of  track 1 \\
     6 & Pseudorapidity $\eta$ of  track 2 \\
     7 & Number of other tracks crossing the beamline between  tracks 1 and 2\\
     \bottomrule
  \end{tabular}
  \caption{Input features for the edge classification BDT. \label{tab:BDTvars}}
\end{table}

\begin{figure}[htbp]
  \centering
    \includegraphics[width=0.5\textwidth]{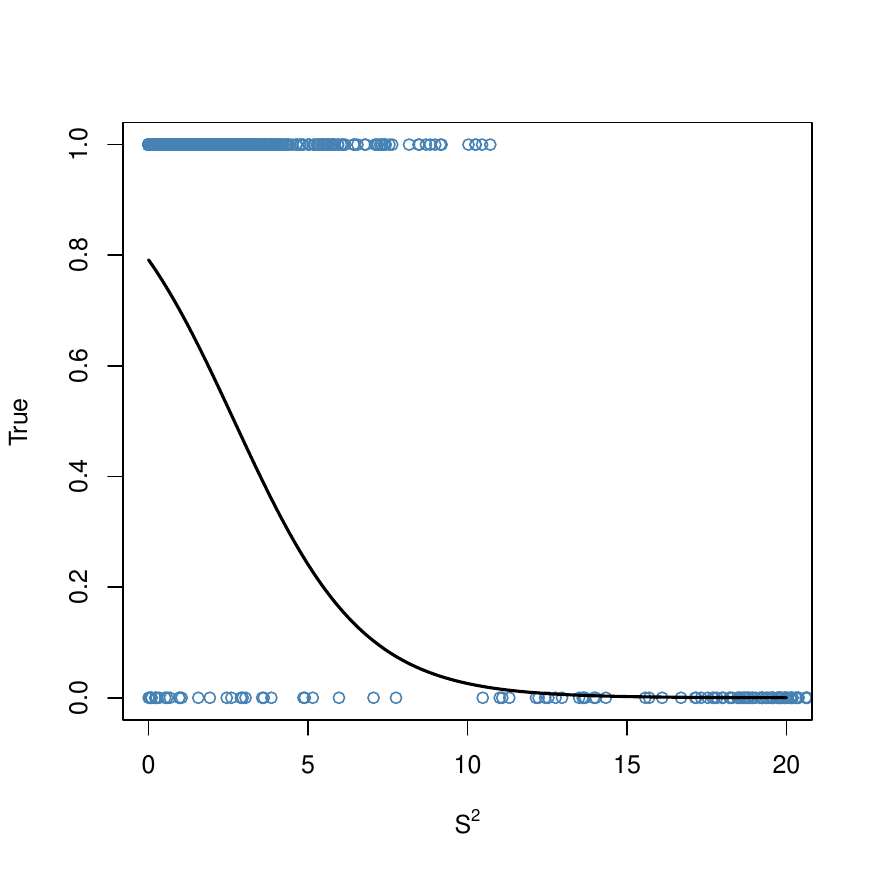} 
  \caption{Example one-variable logistic regression for true (top circles) and 
         fake (bottom circles) edges using the squared track--track distance significance $S^2$.}
  \label{fig:EdgeRegr1}
\end{figure}

Edge weights can also be assigned by using the logistic regression 
$p= e^{z}/(e^z+1)$, where
{$z=\beta_0+\sum_{i=1}^{n}\beta_i x_i$} and $x_i$ are explanatory variables. The negative inverse of the logistic function, $\mathrm{logit}(p)=\log[p/(1-p)]$, provides the necessary edge weight behaviour. Edges that need to be removed receive negative weights, and those that need to be preserved receive positive weights. The intercept value $\beta_0$ is defined by the ratio between the amount of true and fake edges used for training, which can be linked to a prior probability of a given edge being true or fake.  In the current problem, the prior probability depends on the true vertex density and cannot be defined unambiguously, e.g.~it depends on the range of the track--track distance significance $S$, see above. Therefore, the value of the intercept $\beta_0$ in this approach can be modified in some range to achieve an over- or undersegmentation in order to validate its optimality. This will be further discussed in \cref{subsec:stability}. 
A one-dimensional regression is tested in this paper, using variable (1) from \cref{tab:BDTvars}.
The logistic regression for the edge weight calculation is illustrated in \cref{fig:EdgeRegr1}.

The usage of the track--track distance significance for the graph partitioning does not guarantee the compactness --- the limited size and the absence of overlap with other clusters --- of the obtained track clusters in the Cartesian space, which may be beneficial when the vertex density is large. The compactness requirement can be imposed using the LMC constraint mechanism. Some edges in the connectivity graph can be additionally labelled as ``have to be cut'', based on a priori information, different from the edge probability itself. To make track clusters more compact, we can constrain the edges to be cut if the corresponding Cartesian track--track distance is larger than some scale. In the following, a rather weak requirement of  $|z_i-z_j|<1$ mm will be used, which removes tracks with very large errors, see~\cref{fig:TruthVertex}b.  In addition to improving the quality of the solution, the constraint limits the phase space of possible solutions, and this leads to a significant algorithm speedup.  

\section{Performance metrics}
\label{sec:metrics}

 For a quantitative assessment of the performance of the vertex-finding algorithm, one or several metrics are to be established. To compare the performance of the clustering algorithms in, e.g., image segmentation problems, metrics are usually employed, which are based on the comparison of the obtained assignment of the participating objects to clusters with the truth. One example of such a metric is the Variation of Information (VI) proposed in Reference~\cite{MEILA2007873}. The VI metric calculates the degree of compatibility of a clustering $C$ with another clustering $C^{\prime}$ as \begin{equation}
 VI(C,C^{\prime})=H(C)+H(C^{\prime})-2 \cdot I(C,C^{\prime})\end{equation} with 
 \begin{equation}H(C)=-\sum^K_{k=1}P(k) \cdot \log(P(k))~~ {\rm and} ~~I(C,C^{\prime})=\sum^K_{k=1}\sum^{K^{\prime}} _{k^{\prime}=1} P(k,k^{\prime}) \cdot \log\left(\frac{P(k,k^{\prime})}{P(k)P(k^{\prime})}\right)\, .
 \end{equation}Here $P(k)=n_k/N$,~$P(k,k^{\prime})=|C_k \cap C^{\prime}_{k^{\prime}}|/N$, $n_k$ is the number of nodes in the cluster $C_k$, $N$ is the total number of nodes in the graph, and $K$ and $K^{\prime}$ are the number of elements in $C$ and $C^{\prime}$, respectively. In our case, the VI metric can be used to compare the truth track-to-vertex assignment with the obtained clustering solution.
 When the obtained set of track clusters and the track-to-cluster assignment reproduce the truth exactly, $VI$ vanishes. Consequently, smaller VI values correspond to more truth-like (and therefore better) clustering solutions. 
 
 Another track-to-cluster-based metric, which is investigated in the following, is the Silhouette~\cite{Silhouette_1987} score 
 \begin{equation}
 s(i)=\frac{b(i)-a(i)}{\max\{a(i),b(i)\}}
 \end{equation}
 with 
 \begin{equation}
 a(i)=\frac{1}{n_k-1}\sum_{j,~i \ne j}^{C_k}d(i,j) ~~{\rm and }~~ b(i)=\min_{C_{k^{\prime}} \ne C_k}\frac{1}{n_{k'}} \sum_{j}^{C_{k^{\prime}}}d(i,j)
 \end{equation}
 for node $i$ in cluster $C_k$. Here $d(i,j)$ is a distance between nodes $i$ and $j$. In this study, we use the Cartesian distance between tracks and average over all tracks {\it silhouette} value \expect{s(i)} as a quality estimator of the clustering solution. The {\it silhouette} value is limited $-1<s(i)<1$, larger  values corresponding to more compact clusters, better separated from each other.
 
 Several other metrics for the assessment of the clustering performance can be found in Reference~\cite{MEILA2007873}. These metrics are expected to encounter problems in the present case due to the truth track overlap, as explained in~\cref{sec:truthfeatures}. Tracks are assigned most probably to the wrong cluster by any partitioning algorithm if placed in between tracks from other clusters by mismeasurement. This phenomenon inevitably reduces the accuracy of any track-to-cluster-based metrics. 
 Nevertheless, at least the clustering of the well-measured tracks should reproduce the truth closely, which the track-to-cluster metrics can still be sensitive to.
 
 As the metric accuracy is compromised by the presence of tracks with large measurement errors, it might be useful to downscale the contribution of such tracks to the metric. For the VI metric this can be achieved by weighting every track with $\sigma^{-2}$ in the metric calculations, namely $n_k \rightarrow {\mathscr{n}}_k{=}\sum_{i=1}^k\frac{1}{\sigma_i^2}$, $N\rightarrow \mathscr{N}{=}\sum_{i=1}^N\frac{1}{\sigma_i^2}$, etc. For the Silhouette metric the Cartesian distance between two tracks can be replaced by its significance $d(i,j)=S_{ij}$. The weighted versions of the VI and Silhouette metric will be used in the following, along with the original versions.
 
\begin{figure}[htb]
  \centering
    \includegraphics[width=0.6\textwidth]{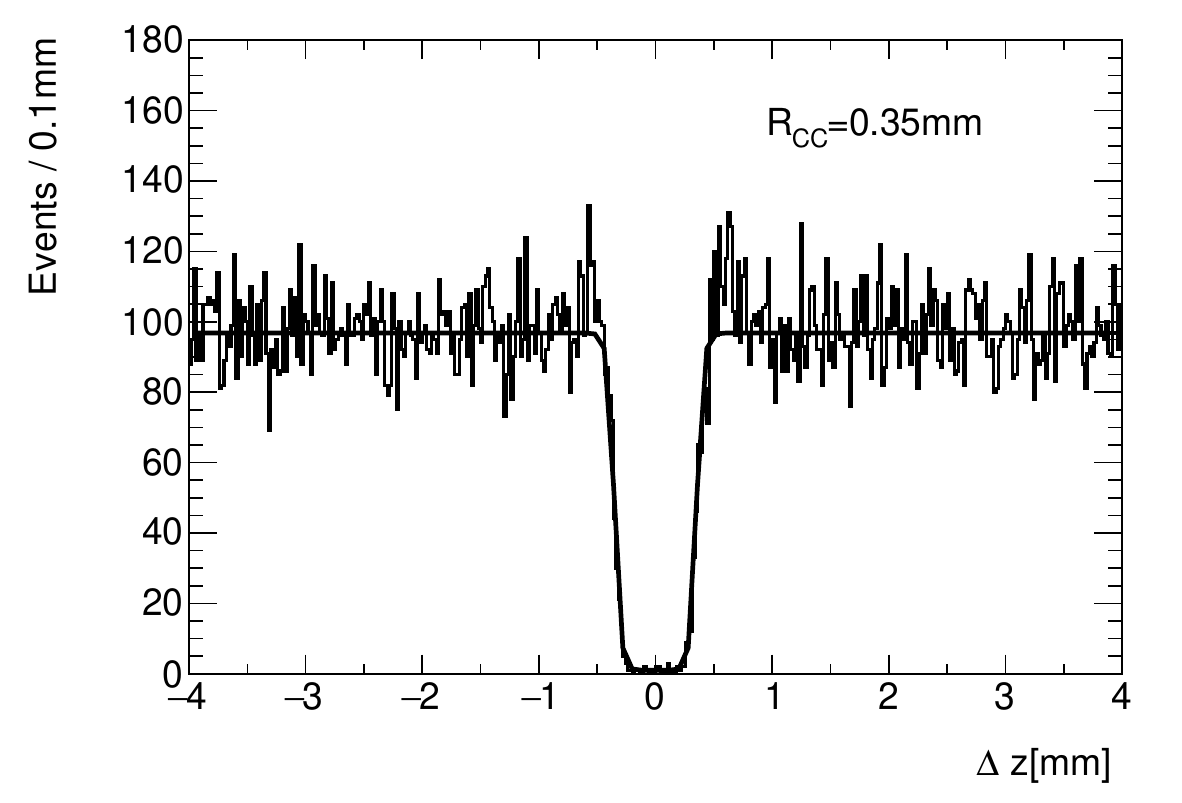} 
  \caption{ Example of a fit to the cluster--cluster distance to determine the resolution. The used fitting function is $a/\{1+\exp [b\cdot(R_{cc}-|x|)]\}+c$ where {\it a, b, c} are free fitting parameters and $R_{cc}$ is the cluster--cluster resolution, defined as the half-width at the half-depth of the dip in the centre of the cluster--cluster weighted centre distances, averaged over all clusters. \label{fig:ccdistfit}}
\end{figure}

 The number of reconstructed clusters and the weighted average positions of these clusters, dominated by the well-measured tracks, are mostly decoupled from the details of the track-to-cluster assignment. The number of clusters can be directly used as a metric (up to the possible presence of fake clusters), but a Cartesian distance-based metric is not straightforward. One may try to introduce such a metric exploiting the cluster--cluster resolution $R_{cc}$, i.e.~the minimal distance between two reconstructed clusters, see \cref{fig:ccdistfit}. The {\it good, merged, bad} cluster categories could be defined based on whether the cluster--truth vertex distance is smaller or larger than $R_{cc}$. Such cluster categories could be used to compare various clustering solutions. But this categorisation explicitly depends on $R_{cc}$, which itself depends on the clustering algorithm. To avoid such circular dependence, a scale-independent Cartesian distance-based metric is needed.

 To construct such a metric, we propose the following procedure. Every reconstructable truth vertex is linked to the closest reconstructed cluster in the Cartesian space that has 2 or more assigned tracks. Thus, a list of linked reconstructed clusters is obtained. Then, every reconstructed cluster is classified depending on how many times it enters into this list. If a cluster enters this list only once, there is just a single truth vertex referencing this cluster. Therefore it can be called {\it unique}, which means that a truth vertex is unambiguously reconstructed as a cluster. If a cluster enters several times into the list, it is referenced by several truth vertices, and therefore it combines tracks from these vertices: this cluster can be called {\it merged}. Also, some clusters may not appear in this list at all: such clusters are not referenced by any truth vertex and are thus {\it fake}. 
The total number of obtained clusters and their classification as {\it unique, merged, fake} are scale-independent and can be used as a metric to compare various clustering options.

\section{Results}
\label{sec:results}

\subsection{LHC Run-2 13 TeV data}
\label{subsec:pileup63}

\begin{figure}[htb]
  \centering
    \includegraphics[width=0.98\textwidth]{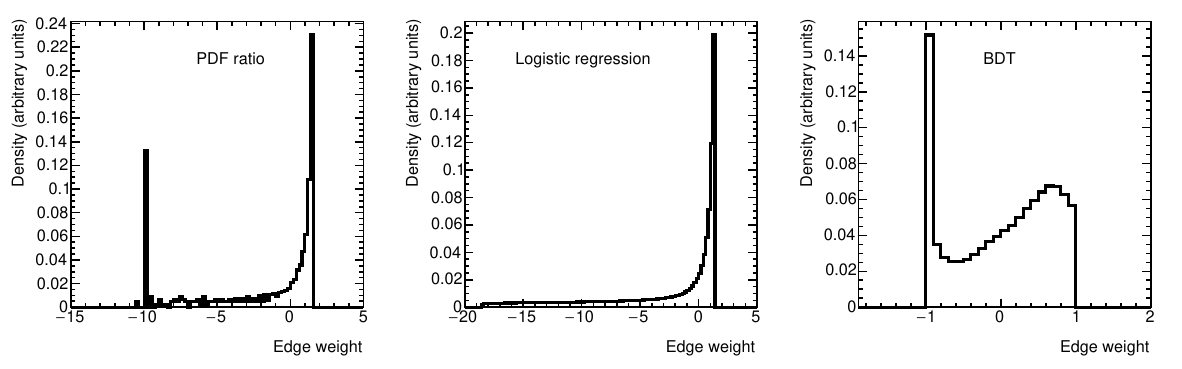} 
  \caption{ Typical edge weight distributions for various edge 
  labelling options. \label{fig:WGTs_dz35_pu63}}
\end{figure}

First, the LMC clustering algorithm is tested with simulated DELPHES data at a collision energy of 13~\TeV, with pileup $\expect{\mu} = 63$ and $\sigma_z  =  35$ mm. These parameters are chosen to provide simulated data close to the actual data collected by the ATLAS detector in Run 2. Due to the very small transverse width of the ATLAS proton--proton interaction region (${\le}\,10\,\mu$m), this width is neglected in the simulation, i.e.~$\sigma_{x,y}  =  0$. Edge-weight distributions for various edge-labelling approaches on these data are shown in \cref{fig:WGTs_dz35_pu63}. The performance of the LMC algorithm on these data is shown in Table~\ref{tab:ClstRes63}. The rows labelled ``cnst'' in these tables provide performance estimation with the applied constraints $|z_i-z_j|<1$ mm, while the ``base'' rows describe the baseline algorithm performance without constraints.

As mentioned above, the primary-vertex reconstruction problem is effectively one-dimensional, all tracks and vertices being located along the beam-line. In an ideal case, the reconstructed clusters are strictly separated on this line, i.e.~they do not contain tracks assigned to another cluster within their geometrical borders. However, the LMP clustering is based on the distance significance and not on the Cartesian distance itself. Thus, a track may be assigned to another cluster nearby. The column \Ntrkwrong in \cref{tab:ClstRes63} shows the fraction of tracks in an event assigned to a cluster, but entirely surrounded by tracks from another cluster along the line. This number is an estimator of the degree of cluster overlap in the obtained solution. The relevant truth track overlap fraction in the same events can be found in \cref{tab:DataSamples} for comparison. The truth track overlap fraction characterises the initial complexity of the event, while the cluster overlap fraction characterises the quality of the obtained solution. 
In addition, \cref{tab:AllEffData} in the Appendix gives the number of isolated nodes (tracks) reported by the LMC clustering algorithm. These non-assigned tracks do not represent the one-track truth vertices, considered non-reconstructable without a priori information, but rather reflect the clustering problems. 

\begin{table}[htb]
  \centering
   \begin{tabular}{cc|cccccccccc}
    \toprule
\multicolumn{2}{c|}{\textsf{\small Edge weight}} & \textsf{\small VI} & \textsf{\small VI} & \textsf{\small Silhouette} & \textsf{\small Silhouette} & \textsf{\small Unique} & \textsf{\small Merged} & \textsf{\small Fake} & \Ntrkwrong & \textsf{\small CPU}\\
\multicolumn{2}{c|}{}  &  & \textsf{\small weighted} & & \textsf{\small weighted} & & & & & \\ 
\midrule
\multirow{2}{*}{\textsf{\small PDF ratio}} & \textsf{\small base} & 0.839 & 0.407 & 0.615 & 0.646 & 33.3 & 8.2 & 2.4 & 15\%&0.25s\\
         & \textsf{\small cnst} & 0.782 & 0.362 & 0.649 & 0.660 & 33.9 & 7.9 & 2.3 & 8\%&0.18s\\
\midrule
\multirow{2}{*}{\textsf{\small Regression}} &\textsf{\small base} & 0.860 & 0.416 & 0.589 & 0.623 & 34.7 & 7.6 & 4.1 & 14\% & 0.27s\\
          &\textsf{\small cnst} & 0.829 & 0.387 & 0.614 & 0.633 & 35.0 & 7.5 & 3.9 & 8\% & 0.18s\\
\midrule
  \multirow{2}{*}{\textsf{\small BDT}}    & \textsf{\small base} & 0.945 & 0.399 & 0.478 & 0.230 & 35.0 & 7.5 & 7.1 & 5\%&0.23s\\
         & \textsf{\small cnst} & 0.937 & 0.377 & 0.487 & 0.234 & 35.2 & 7.4 & 7.0 & 4\%&0.14s\\
\bottomrule
  \end{tabular}
  \caption{ LMC performance for the collision energy \SI{13}{TeV}, pileup 63 and interaction region width $\sigma_z = \SI{35}{mm}$. These simulation parameters are chosen to match the full ATLAS simulation for Run 2 results used for comparison. The column \Ntrkwrong shows the fraction of tracks wrongly associated by the clustering algorithm, which can be compared to the truth track overlap fraction of 20\% (\cref{tab:TruthOverlap}).  \label{tab:ClstRes63}}
\end{table}

  The PDF ratio and the regression-based edge weight assignment result in approximately equal clustering performance. The BDT-based edge weight assignment leads to a significantly worse Silhouette metric value, a smaller value of the cluster overlap and a larger amount of fake clusters. As expected, the weighted versions of the VI and Silhouette metrics have significantly better values than the standard ones due to downscaling of the noise. Using constraints uniformly improves all quality estimators and provides $\sim30$\% CPU reduction.  

\begin{figure}[htb]
  \centering
    \includegraphics[width=0.98\textwidth]{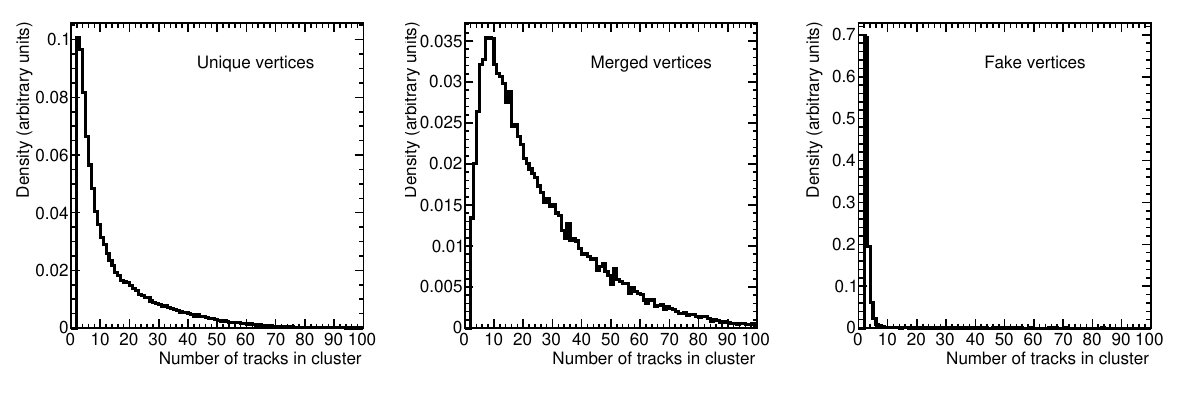} 
  \caption{ Number of tracks in a cluster for the {\it unique, merged} and {\it fake} cluster categories. The distributions are obtained for pileup $\expect{\mu}=63$ data using a one-variable logistic regression for the edge weight assignment.
  \label{fig:UniqueMergedFake}}
\end{figure}

In total, $70$\% of the reconstructable truth vertices are reconstructed as {\it unique} clusters, while the remaining 30\% (i.e.~15) truth vertices are squeezed into $7.5$ {\it merged} vertices. The amount of fake clusters is in the range of 5--15\%. The number of tracks in the different cluster categories is presented in \cref{fig:UniqueMergedFake}. The number of tracks in the {\it unique} clusters is close to the track amount in the truth vertices, see \cref{fig:TruthVertex}, while the {\it merged} clusters contain much more tracks.  Finally, {\it fake} clusters have a very small number of tracks.

\subsection{High-Luminosity LHC 14 TeV data}
\label{subsec:pileup250}

The High Luminosity LHC (HL-LHC) project foresees a significant increase in interaction rates to collect significantly more data and thus increase the sensitivity for new physics. The exact parameters of the upgraded HL-LHC are not yet final; pileup values of 150, 200, and 250, and an interaction region width of $\sigma_z = \SI{42}{mm}$~\cite{HLLHCperf} are considered in this paper. These options result in an increase in the density of pileup interaction vertices up to a factor of 4, as compared to the current LHC parameters. The truth track overlap fraction rises from 20\% to 66\%, see \cref{tab:DataSamples}. It is interesting to check the performance of the LMC problem formulation in such extreme conditions.

For this test, the same PDF ratio and logistic regression function are used for the edge weight calculation, while the BDT classification is retrained using $\mu=150,200,250$ data. Results for nominal PDF ratio and logistic regression-based edge weight calculation functions are shown in Tables~\ref{tab:ClstRes150}, \ref{tab:ClstRes200}, and \ref{tab:ClstRes250}.

\begin{table}[htb]
  \centering
   \begin{tabular}{cc|cccccccccc}
    \toprule
\multicolumn{2}{c|}{\textsf{\small Edge weight}} & \textsf{\small VI} & \textsf{\small VI} & \textsf{\small Silhouette} & \textsf{\small Silhouette} & \textsf{\small Unique} & \textsf{\small Merged} & \textsf{\small Fake} & \Ntrkwrong & \textsf{\small CPU}\\
\multicolumn{2}{c|}{}  &  & \textsf{\small weighted} & & \textsf{\small weighted} & & & & &\\ 
\midrule
 \multirow{2}{*}{\textsf{\small PDF ratio}} & \textsf{\small base} & 1.318 & 0.690 & 0.535 & 0.577 & 57.7 & 27.4 & 4.8 & 28\% & 1.1s\\
         & \textsf{\small cnst} & 1.211 & 0.612 & 0.581 & 0.609 & 59.4 & 26.9 & 4.1 & 14\% & 0.42s\\
\midrule
\multirow{2}{*}{\textsf{\small Regression}} &\textsf{\small base} & 1.316 & 0.682 & 0.514 & 0.559 & 63.0 & 25.6 & 8.8 & 26\% & 0.73s\\
          &\textsf{\small cnst} & 1.259 & 0.634 & 0.546 & 0.582 & 63.6 & 25.4 & 8.2 & 14\% &0.50s\\
\midrule
 \multirow{2}{*}{\textsf{\small BDT}}    & base & 1.303 & 0.658 & 0.394 & 0.146 & 61.8 & 25.9 & 13 & 9\% & 0.96s\\
         & \textsf{\small cnst} & 1.275 & 0.616 & 0.409 & 0.155 & 62.8 & 25.6 & 12 & 7\% & 0.43s\\
\bottomrule
  \end{tabular}
  \caption{ LMC performance for pileup $\mu=150$ in an HL-LHC environment with  collision energy \SI{14}{TeV} and interaction region size $\sigma_z = \SI{42}{mm}$.
  The column \Ntrkwrong shows the fraction of the tracks, wrongly associated by the clustering algorithm, which can be compared to the truth track overlap fraction 41\% (\cref{tab:TruthOverlap}).
  \label{tab:ClstRes150}}
\end{table}

\begin{table}[htb]
  \centering
   \begin{tabular}{cc|cccccccccc}
    \toprule
\multicolumn{2}{c|}{\textsf{\small Edge weight}} & \textsf{\small VI} & \textsf{\small VI} & \textsf{\small Silhouette} & \textsf{\small Silhouette} & \textsf{\small Unique} & \textsf{\small Merged} & \textsf{\small Fake} & \Ntrkwrong & \textsf{\small CPU}\\
\multicolumn{2}{c|}{}  &  & \textsf{\small weighted} & & \textsf{\small weighted} & & & & & \\ 
\midrule
\multirow{2}{*}{\textsf{\small PDF ratio}} & \textsf{\small base} & 1.574 & 0.852 & 0.500 & 0.546 & 64.3 & 40.3 & 5.7 & 36\% &2.3s\\
         & \textsf{\small cnst} & 1.441 & 0.756 & 0.552 & 0.586 & 66.6 & 39.8 & 4.8 & 18\% &0.69s\\
\midrule
\multirow{2}{*}{\textsf{\small Regression}} &\textsf{\small base} & 1.546 & 0.825 & 0.492 & 0.539 & 70.3 & 38.6 & 9.0 & 32\% &2.4s\\
          &\textsf{\small cnst} & 1.470 & 0.765 & 0.529 & 0.568 & 71.0 & 38.4 & 8.1 & 18\%&0.69s\\
\midrule
\multirow{2}{*}{\textsf{\small BDT}}   & \textsf{\small base} & 1.512 & 0.805 & 0.312 & 0.040 & 69.9 & 38.6 & 15.6 & 13\%&1.8s\\
         & \textsf{\small cnst} & 1.479 & 0.755 & 0.332 & 0.051 & 71.3 & 38.2 & 15.0 & 7\%&0.66s\\
\bottomrule
  \end{tabular}
  \caption{ LMC performance for pileup $\mu=200$ in an HL-LHC environment with  collision energy \SI{14}{TeV} and  interaction region size $\sigma_z = \SI{42}{mm}$. 
   The column \Ntrkwrong shows the fraction of the tracks, wrongly associated by the clustering algorithm, which can be compared to the truth track overlap fraction 53\% (\cref{tab:TruthOverlap}).
  \label{tab:ClstRes200}}
\end{table}

\begin{table}[htb]
  \centering
   \begin{tabular}{cc|cccccccccc}
    \toprule
\multicolumn{2}{c|}{\textsf{\small Edge weight}} & \textsf{\small VI} & \textsf{\small VI} & \textsf{\small Silhouette} & \textsf{\small Silhouette} & \textsf{\small Unique} & \textsf{\small Merged} & \textsf{\small Fake} & \Ntrkwrong & \textsf{\small CPU}\\
\multicolumn{2}{c|}{}  &  & \textsf{\small weighted} & & \textsf{\small weighted} & & & & & \\ 
\midrule
\multirow{2}{*}{\textsf{\small PDF ratio}} & \textsf{\small base} 
  & 1.782 & 0.990 & 0.477 & 0.526 & 68.7 & 53.2 & 6.4 & 42\% & 3.0s\\
  & \textsf{\small cnst} 
  & 1.638 & 0.887 & 0.531 & 0.569 & 71.0 & 52.7 & 5.3 & 21\% & 1.7s\\
\midrule
\multirow{2}{*}{\textsf{\small Regression}} &\textsf{\small base} 
  & 1.753 & 0.961 & 0.467 & 0.517 & 77.1 & 51.2 & 11. & 38\% & 3.2s\\
  &\textsf{\small cnst} 
  & 1.672 & 0.895 & 0.505 & 0.547 & 77.8 & 51.1 & 9.9 & 21\% & 1.7s\\
\midrule
 \multirow{2}{*}{\textsf{\small BDT}} & \textsf{\small base} 
  & 1.691 & 0.941 & 0.307 & 0.040 & 72.8 & 52.4 & 15. & 12\% & 3.0s\\
  &\textsf{\small cnst} 
  & 1.651 & 0.882 & 0.330 & 0.055 & 74.5 & 52.0 & 14. & 9\% & 1.2s\\
\bottomrule
  \end{tabular}
  \caption{ LMC performance for pileup $\mu=250$ in an HL-LHC environment with collision energy \SI{14}{TeV} and interaction region size $\sigma_z = \SI{42}{mm}$. 
  The column \Ntrkwrong shows the fraction of the tracks, wrongly associated by the clustering algorithm, which can be compared to the truth track overlap fraction 66\% (\cref{tab:TruthOverlap}).
  \label{tab:ClstRes250}}
\end{table}

Similarly to the $\mu=63$ results, the BDT-based edge weight assignment leads to a significantly worse Silhouette metric value, a much smaller value of the cluster overlap and a larger number of fake clusters, while the PDF ratio and regression-based edge weight calculation approaches provide similar performances. The weighted versions of the VI and Silhouette metrics have significantly better values than the standard ones due to downscaling of the noise. The use of constraints significantly improves all quality estimators and provides $\sim30$\% CPU reduction.  

The number of unambiguously reconstructed {\it unique} clusters is 53\% (44\%, 37\%) out of the total amount of the reconstructable truth vertices for the pileup $\mu=150$ (200, 250). The remaining 56 (90, 125) reconstructable truth vertices are clustered into 25 (40, 52) {\it merged} clusters. The correctness of representation of the initial truth vertices by {\it merged} clusters is not granted. Truth vertices with a large number of tracks might ``absorb'' vertices with a small number of tracks.

\subsection{LMC performance adjustment}
\label{subsec:stability}

\begin{figure}[htb]
  \centering
    \includegraphics[width=0.9\textwidth]{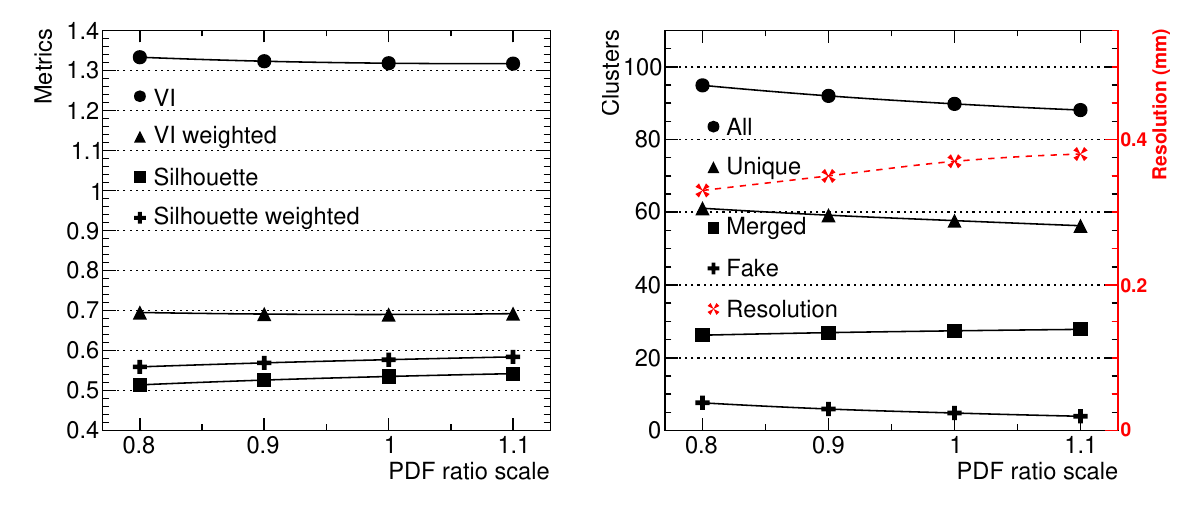} 
  \caption{ PDF ratio cost-based clustering results as a function of the applied scaling.
  \label{fig:RatioScale}}
\end{figure}

\begin{figure}[ht]
  \centering
    \includegraphics[width=0.9\textwidth]{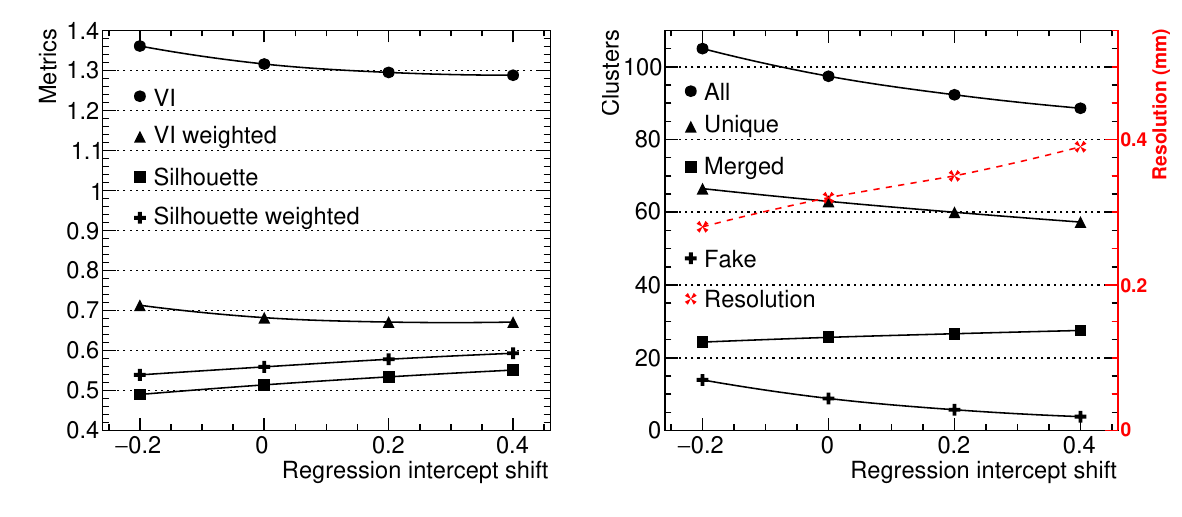} 
  \caption{ Logistic regression cost-based clustering results as a function of the logistic regression intercept term shift.
  \label{fig:RegrShift}}
\end{figure}

As can be seen from Tables~\ref{tab:ClstRes63}--\ref{tab:ClstRes250},
different edge weight assignment approaches lead to non-coinciding clustering results. For a practical application of the LMC approach for primary vertex finding in the LHC experiments, it is important to verify whether a unique optimal clustering solution exists in this problem and, if so, whether the different LMC cost functions can be tuned to provide the same clustering. 
As explained in \cref{sec:wgtedge}, parameters of the PDF ratio and regression function for the edge weights can be modified to enforce under- or over-segmentation.The {\it PDF} ratio function can be scaled up and down. In the logistic regression function, the intercept term can be shifted by a constant. The cost function modifications are tried on the $\mu=150$ data. The obtained clustering results are shown in \cref{fig:RatioScale} and \cref{fig:RegrShift}.

In the performed test, the exploited metrics change monotonically depending on the scale factor for the PDF ratio  and  the intercept shift for the linear regression function. It doesn't seem possible to adjust the PDF ratio and logistic regression parameters so that both approaches provide exactly the same clustering performances in all used metrics.  In addition, the BDTG-based Silhouette and Silhouette weighted metrics results (see Table~\ref{tab:ClstRes150}) are not reproducible by any modification of the PDF ratio and logistic regression cost functions. However, the overall variations of the clustering results remain limited, which means that the LMC approach performance stays close to optimal in the full scanned parameter range.

 To conclude, the cost function modification test doesn't demonstrate the presence of an evident unique globally optimal clustering solution for the problem in consideration. Three used edge weight assignment strategies provide different clustering results, which can be additionally changed by simple modification of the cost functions. Therefore, for a practical application as a primary vertex finder, an exact LMC formulation should be chosen based on desired physics requirements, e.g. minimal amount of fake vertices or best vertex--vertex resolution, disregarding the clustering metrics.

\subsection{Influence of tracks with large measurement errors}
\label{subsec:noise}

\begin{table}[htb]
  \centering
   \begin{tabular}{ccc}
    \toprule
\textsf{\small Track error cut}& \Ntrk & \textsf{\small Truth track overlap}\\
\midrule
- & 1674 & 41\%  \\
0.8 & 1540 & 31\%  \\
0.6 & 1444 & 27\% \\
0.4 & 1283 & 22\%  \\
\bottomrule
  \end{tabular}
  \caption{ Number of selected tracks and the truth track overlap fraction as a function of the track error cut for $\mu=150$ data.
  \label{tab:zcut150}}
\end{table}

\begin{figure}[htb]
  \centering
    \includegraphics[width=0.9\textwidth]{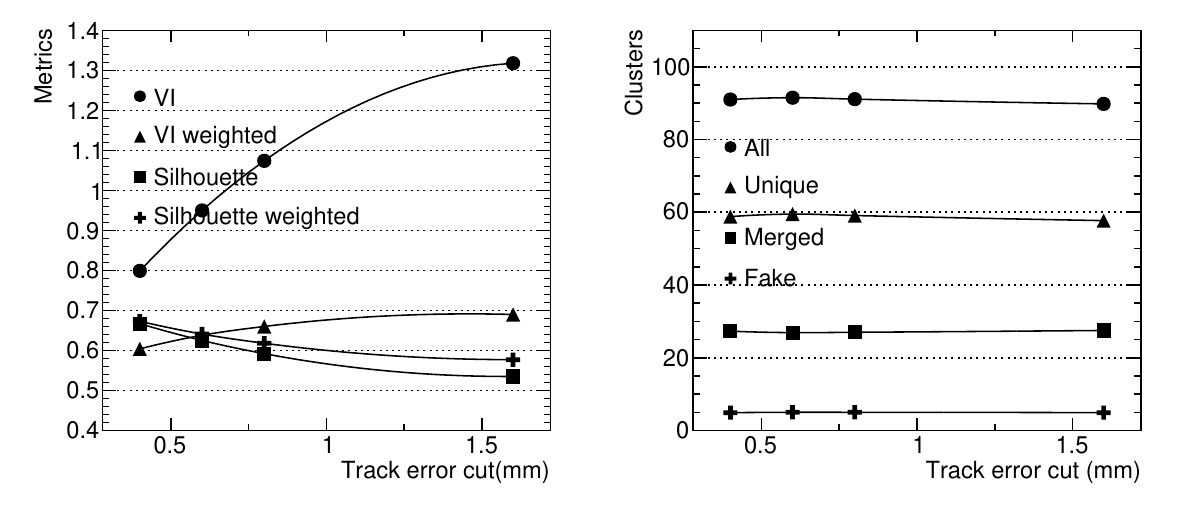} 
  \caption{ PDF ratio cost-based clustering results as a function of the applied track error cut for the $\mu=150$ data.
  \label{fig:badtrk_ratio}}
\end{figure}

\begin{figure}[htb]
  \centering
    \includegraphics[width=0.9\textwidth]{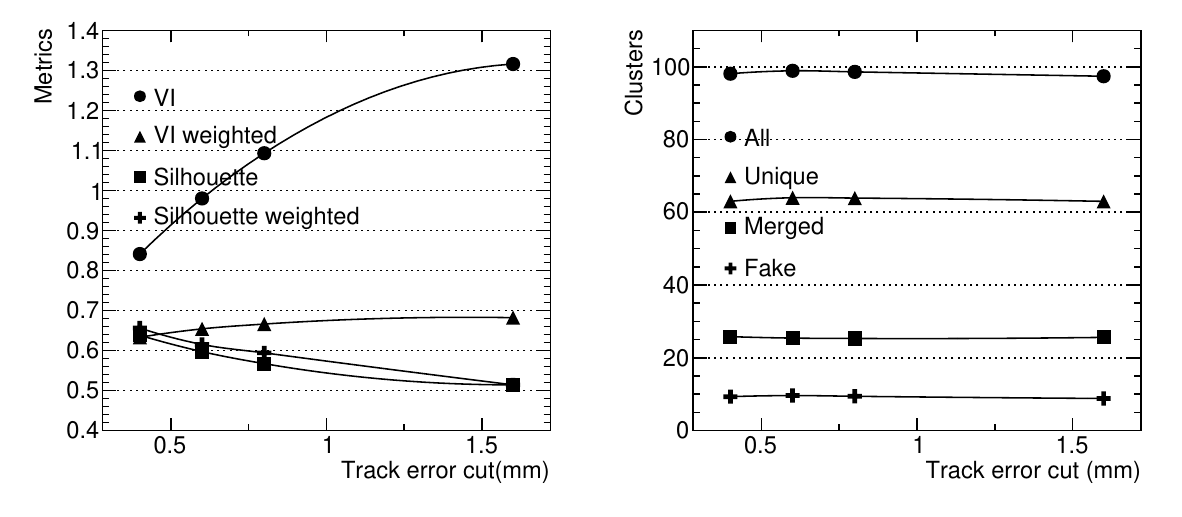} 
  \caption{ Logistic regression cost-based clustering results as a function of the applied track error cut for the $\mu=150$ data.
  \label{fig:badtrk_regr}}
\end{figure}

As the initial truth track overlap in an event is caused by the track position mismeasurement, the overlap degree can be reduced by removing the badly measured tracks by cutting on the track measurement error shown in \cref{fig:TruthVertex}b. A moderate decrease in the total amount of tracks due to this rejection should not significantly affect the overall clustering efficiency as the total amount of tracks per truth vertex is big enough, see \cref{fig:TruthVertex}a. The reduction of the amount of selected tracks and the truth track overlap fraction due to strongly mismeasured track removal is shown in \cref{tab:zcut150}. The results of the clustering are shown in \cref{fig:badtrk_ratio} for the PDF ratio cost function and in  \cref{fig:badtrk_regr} for the nominal logistic regression cost function.

The distance-based metric demonstrates very small changes in the clustering results in a wide range of the badly measured track admixture and, correspondingly, the initial degree of the vertex overlap. One may conclude that the amount of clusters identified by the LMC algorithm is largely defined by the tracks with small measurement errors and, therefore, is stable with respect to significant track noise admixture. Redistribution of the tracks with big errors over the obtained clusters doesn't change their amount but evidently strongly affects all track counting-based clustering metrics. The track weighting does mitigate this effect for the VI metric, its weighted version is practically independent of the track noise admixture. Surprisingly, the Silhouette metric is only weakly sensitive to this noise.

\subsection{Comparison with the existing approaches}
\label{subsec:comparison}

\begin{figure}[htb]
  \centering
   \begin{tabular}{cc}
    \includegraphics[width=0.49\textwidth]{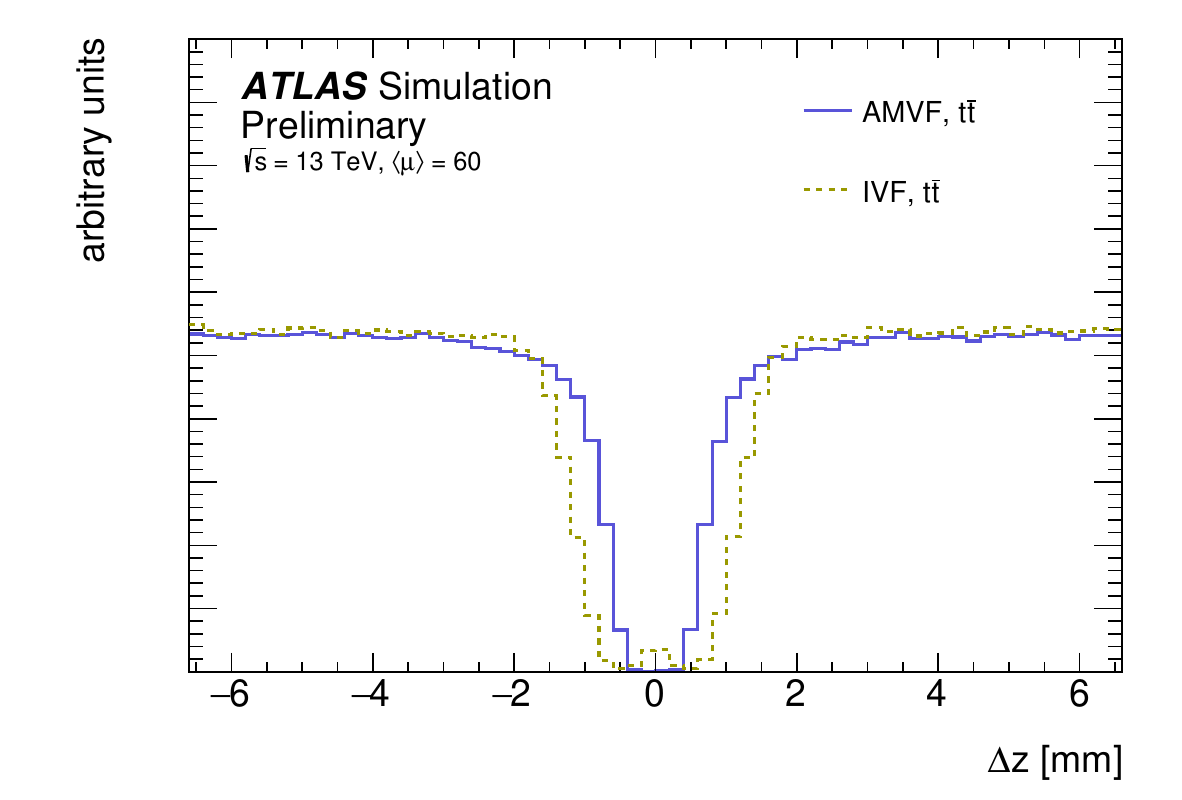} &
    \includegraphics[width=0.46\textwidth]{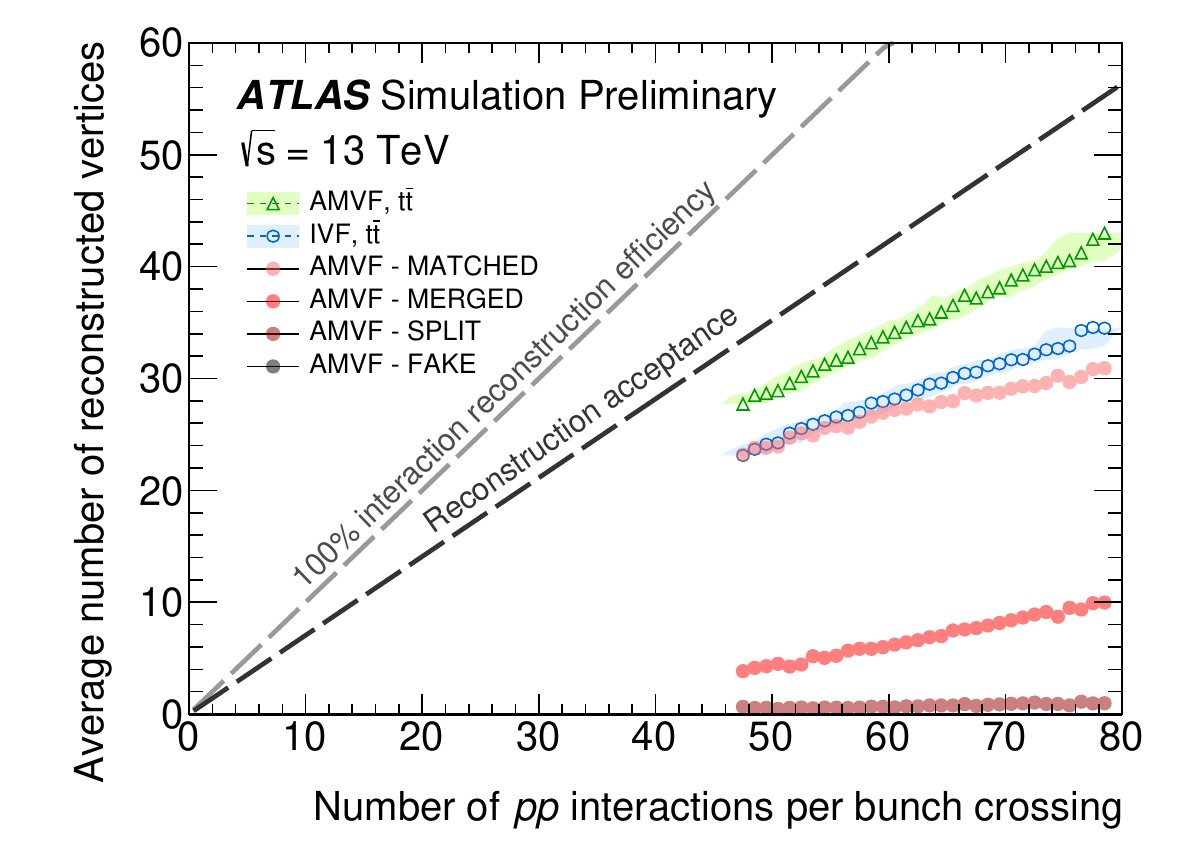} \\  
    \includegraphics[width=0.49\textwidth]{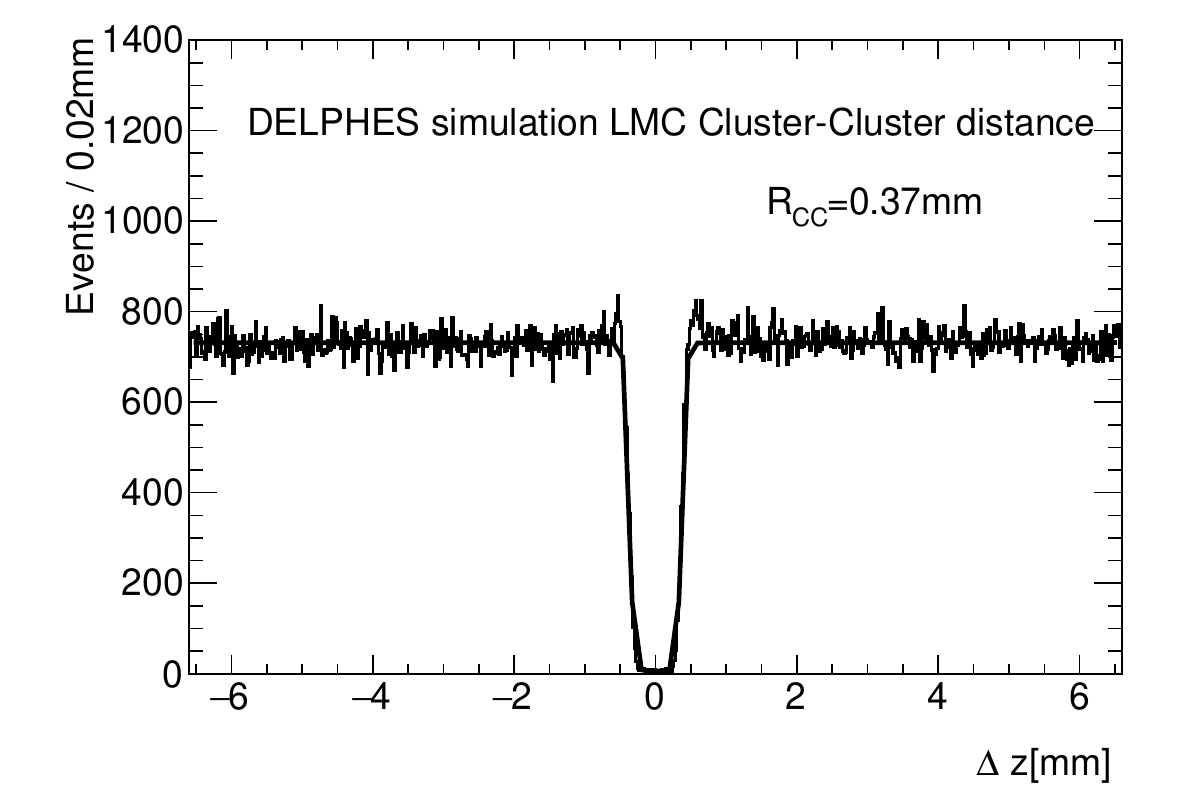} &
    \includegraphics[width=0.46\textwidth]{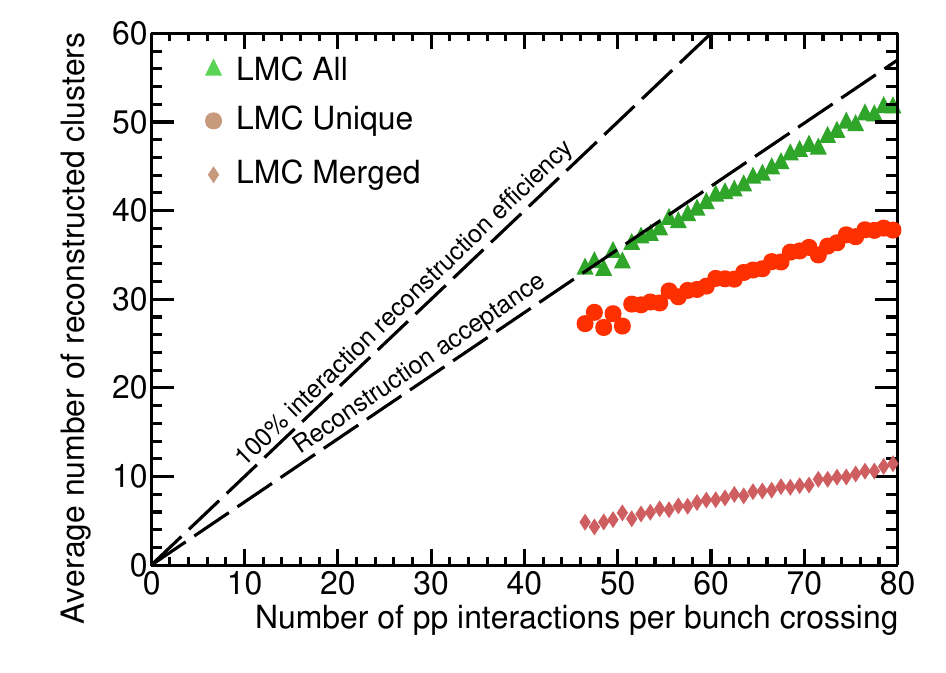}\\
  \end{tabular}
  \caption{The vertex--vertex resolution and the number of reconstructed vertices as a function of the number of $pp$ interactions for typical ATLAS data. The upper plots are obtained with the ATLAS baseline AMVF~\cite{PUB-2019-015} and IVF~\cite{IVF} algorithms.
  The bottom plots are obtained using the LMC algorithm with the PDF ratio-based edge weight assignment on DELPHES $\mu=63$ data. The DELPHES $\mu=63$ simulation is specially tuned to match the ATLAS data used in \cite{PUB-2019-015}. The cluster--cluster resolution for the LMC algorithm on the bottom left picture is obtained as described in~\cref{sec:metrics}.  \label{fig:results}}
\end{figure}

 The ATLAS Collaboration used the IVF algorithm~\cite{IVF} to reconstruct the $pp$ collision vertices in \mbox{Run 1} and the AMVF algorithm~\cite{PUB-2019-015} in \mbox{Run 2} and \mbox{Run 3}. Essential characteristics of a primary-vertex reconstruction algorithm are the vertex--vertex resolution and the number of reconstructed vertices as a function of the number of $pp$ interactions. The upper plots in \cref{fig:results} present the corresponding distributions for typical ATLAS data for the AMVF and IVF algorithms. The bottom plots show the same distributions provided by the LMC algorithm with the PDF-ratio based edge-weight assignment using DELPHES simulation tuned to the same pileup conditions and track resolutions. The fast DELPHES simulation lacks some features present in the full ATLAS simulation (fakes, inefficiencies, non-Gaussian tails, etc.). However, these features are not expected to change  the primary vertex reconstruction results significantly, in particular, due to the highly efficient ATLAS track reconstruction.

 \cref{fig:results} clearly demonstrates that the LMC algorithm outperforms the ATLAS heuristic algorithms. It provides significantly better vertex--vertex resolution. This naturally leads to a larger amount of {\it Unique/Matched} vertices reconstructed by LMC, while the amount of {\it Merged} vertices remains practically the same.  Routine application of the LMC for the primary vertex reconstruction can provide a significant gain in performance for LHC and future collider experiments.

\section{Conclusion}
\label{sec:conclusions}
 In this work, we have addressed a typical particle physics problem of reconstructing multiple interaction positions in a dense environment, where each interaction is represented by a cluster of tracks. Significant track reconstruction errors lead to a large overlap of truth track clusters, which makes their identification challenging.
Heuristic algorithms are usually used to address this problem.
In contrast, we propose to address this problem through a principled formulation as a minimum-cost lifted multicut problem. We construct several cost functions for the LMC from track--track distances and their significance. We study the performance of the LMC algorithm for different vertex densities, cost functions, constraint usage and varying degree of overlap. To address potential performance problems of existing track counting clustering metrics for strongly overlapped clusters, dedicated metrics are introduced.  

We demonstrate that the LMC approach outperforms the heuristic algorithms in the problem of vertex reconstruction in dense environments in terms of vertex--vertex resolution and vertex reconstruction efficiency. It works up to the highest vertex density expected at the HL-HLC project in spite of the strong truth track overlap reaching $\sim60\%$. Variations of the LMC algorithm parameters and cost functions studied in this work resulted in relatively small variations of the obtained clustering solutions. The developed metrics and the freedom in the choice of the edge-weight assignment strategy allow to fine-tune this algorithm to a specific particle-physics experiment.

\acknowledgments

This work is supported by the German Science Foundation (DFG) through a research grant and a Heisenberg professorship under contracts CR-312/4-1 and CR-312/5-1.

\bibliography{references}
\bibliographystyle{JHEP}

\clearpage

\appendix

\section{Non-clustered tracks and total reconstructed clusters}

In this study, we use four simulated event samples representing realistic proton--proton interactions at the LHC with different energies and luminosities. The total amounts of interaction vertices with one reconstructed track and two and more tracks are shown 
in~\cref{tab:AllEffData}. Due to the track measurement errors, the one-track vertices are difficult to reconstruct correctly without a priori information. Finding two and more track vertices becomes problematic if the vertex--vertex distance is less than the typical track measurement error. Both problems are illustrated in~\cref{tab:AllEffData}, where the amounts of the one-track and multi-track clusters are given for every cost function and event sample. 

\begin{table}[htb]

  \centering
   \begin{tabular}{c|cc|cc|cc|cc}
   \toprule
  & \multicolumn{2}{c|}{ 13 TeV} & \multicolumn{6}{c} {14 TeV}   \\
\cline{2-9}
   & \multicolumn{2}{c|}{$\expect{\mu}=63$} & \multicolumn{2}{c|} {$\expect{\mu}=150$}  &
    \multicolumn{2}{c|} {$\expect{\mu}=200$} & \multicolumn{2}{c} {$\expect{\mu}=250$} \\
\cline{2-9}
 &\Nvrtntrkone& \Nvrtntrkmany& \Nvrtntrkone &\Nvrtntrkmany &\Nvrtntrkone &\Nvrtntrkmany
 &\Nvrtntrkone& \Nvrtntrkmany \\[3pt]
\hline
 Truth & 4 & 50 & 9 & 119 & 12 & 160 & 16 & 199 \\
\hline \hline
 &\Nclntrkone& \Nclntrkmany& \Nclntrkone &\Nclntrkmany &\Nclntrkone &\Nclntrkmany
 &\Nclntrkone& \Nclntrkmany \\[3pt]
\hline
 PDF ratio & 11 & 44 & 19 & 90 & 23 & 110 & 25 & 128 \\
\hline
 Regression & 13 & 46 & 25 & 97 & 27 & 118 &31 & 139 \\
\hline
 BDTG  & 43 & 50 & 77 & 101 & 102 & 124 & 104 & 140 \\
\bottomrule
\end{tabular}
  \caption{ Average numbers of non-clustered tracks and reconstructed clusters obtained by the LMC algorithm with different cost functions as compared to the truth numbers of  single-track and multi-track vertices.  Results are shown for all collision energies and pileup densities. \label{tab:AllEffData}}
\end{table}

  The number of one-track clusters in each case is significantly larger than the truth amount of one-track interaction vertices, especially in the BDT case. They should be thought of as non-clustered tracks, not as reconstructed one-track vertices. The fraction of multi-track clusters found decreases with the interaction vertex density, as expected.  

\end{document}